\documentclass[twocolumn,prd,preprintnumbers,amsmath,amssymb,superscriptaddress,showpacs]{revtex4}

\usepackage{pstricks}
\usepackage{color}
\usepackage{braket}
\usepackage{bm}
\usepackage{booktabs}
\usepackage{graphicx}

\def\expv#1{\left< #1 \right>}
\newcommand{\Slash}[1]{{\ooalign{\hfil/\hfil\crcr$#1$}}}

\newcommand{\be}{\begin{equation}}
\newcommand{\ee}{\end{equation}}
\newcommand{\ba}{\begin{align}}
\newcommand{\ea}{\end{align}}
\newcommand{\nn}{\nonumber}
\DeclareMathOperator{\tr}{\mathrm{Tr}\,}

\newcommand{\Tr}{\mathrm{Tr}}

\newcommand{\PhiM}[2]{ \left( \Phi_{#1} \right)_{#2}}   
\newcommand{\rM}[2]{r_{#1,#2}}    
\newcommand{\tM}[2]{\theta_{#1,#2}}  
\newcommand{\Nf}{N_{f}}
\newcommand{\GM}[1]{G_{#1}}    
\newcommand{\PRTs}[2]{\left( #1  \right)_{ #2}}   
\usepackage[normalem]{ulem}  
\newcommand{\mrm}[1]{\mathrm{#1}}

\newcommand{\equref}[1]{Eq.~(\ref{#1})}
\newcommand{\secref}[1]{Sec.~\ref{#1}}
\newcommand{\figref}[1]{Fig.~\ref{#1}}

\newcommand\comout{\bgroup \color{red} \ULdepth=-.5ex \ULset}
\renewcommand\sout{\bgroup \color{red} \ULdepth=-.5ex \ULset}

\newcommand\Tsout{\bgroup \color{purple} \ULdepth=-.5ex \ULset}

\begin{document}

\preprint{KUNS-2617/YITP-16-46/KEK-CP-342}


\title{
Test for a universal behavior of Dirac eigenvalues in the complex Langevin method}
\author{Terukazu Ichihara}
\email{ichihara.terukazu.45z@st.kyoto-u.ac.jp}
\affiliation{Department of Physics, Kyoto University,
         Kyoto 606-8502, Japan}
\affiliation{Yukawa Institute for Theoretical Physics, Kyoto University,
         Kyoto 606-8502, Japan}

\author{Keitaro Nagata}
\email{knagata@post.kek.jp}
\affiliation{KEK Theory Center, High Energy Accelerator Research Organization,
       1-1 Oho, Tsukuba, Ibaraki 305-0801, Japan  }

\author{Kouji Kashiwa}
\email{kouji.kashiwa@yukawa.kyoto-u.ac.jp}
\affiliation{Yukawa Institute for Theoretical Physics, Kyoto University,
         Kyoto 606-8502, Japan}

\begin{abstract}
We apply the complex Langevin (CL) method to a chiral random
 matrix theory (ChRMT) at non-zero chemical potential and study the nearest neighbor spacing (NNS)
 distribution of the Dirac eigenvalues.
The NNS distribution is extracted using an unfolding procedure for the
 Dirac eigenvalues obtained in the CL method.
For large quark mass, we find that the NNS distribution obeys the Ginibre ensemble as expected.
For small quark mass, the NNS distribution follows the Wigner
 surmise for correct convergence case, 
while it follows the Ginibre ensemble for wrong convergence case.
The Wigner surmise is physically reasonable from the chemical potential
independence of the ChRMT. 
The Ginibre ensemble is known to be favored in a phase quenched QCD at finite chemical potential.
Our result suggests a possibility that the originally universal behavior of the NNS distribution 
is preserved even in the CL method for correct convergence case.
\end{abstract}

\maketitle



\section{Introduction}

The complexification of field theories has attracted
recent attention in the context of solving the sign problem.
The complex Langevin (CL) method
\cite{Parisi:1984cs,Klauder:1983sp,Seiler:2012wz}
and the Lefshetz thimble (LT) method
\cite{Pham:1983,Witten:2010cx,Witten:2010zr}
rely on the complexification of dynamical variables of the system.
Although they are hopeful candidates as a solution for
the sign problem, there are still some controversies over their
feasibility or practicality.
One of the difficulties lies in the lack of knowledge of complexified theories,
which we would like to address in this work.

The Langevin method is based on the stochastic quantization \cite{Parisi:1980ys}.
As for real actions, 
the Langevin method is ensured to produce
correct results in an infinitely large Langevin-time limit, where the
probability distribution of dynamical variables approaches to
the Boltzmann weight as indicated in the
eigenvalue analysis of the Fokker-Planck
equation, e.g. see Refs.~\cite{Namiki:1992,Damgaard:1987rr}.
As for complex actions \cite{Parisi:1984cs,Klauder:1983sp},
observables in the CL method sometimes converge to wrong results.
Recently, Aarts et al.~showed that the CL method can be justified
if some conditions are satisfied~\cite{Aarts:2009uq,Aarts:2011ax}.
Possible reasons of its breakdown are the spread of CL configurations in the
complexified direction in the complex plane of field
variables~\cite{Aarts:2009uq,Aarts:2011ax} 
and singular drift terms of CL equations \cite{Mollgaard:2013qra,Greensite:2014cxa,Nishimura:2015pba}.

A new method, which is referred to as the gauge cooling,
has been proposed to suppress the spread of
configurations in the imaginary direction by using the
complexified gauge invariance~\cite{Seiler:2012wz,Sexty:2013ica,Fodor:2015doa}.
It was shown that the method reproduces correct results of physical
observables in some cases \cite{Seiler:2012wz}.
Later, the gauge cooling method is justified in Ref.~\cite{Nagata:2015uga},
and extended to theories that have a global symmetry with sufficiently large
number of generators~\cite{Nagata:2015ijn}.
In addition, it was found in Ref.~\cite{Mollgaard:2013qra} that the singular drift
problem occurred in a chiral random matrix theory (ChRMT), where complex
eigenvalues of the fermion matrix touch the origin at non-zero chemical potential.
Here the fermion matrix is given as $D+m$ with a Dirac operator $D$.
Zero modes of the fermion matrix cause a singularity in a drift term of the Langevin equation.
Such a singularity originates from the logarithmic term in the effective action, 
which is expected to occur for fermionic theories quite in general.
Mollgaard and Splittorff pointed out that
a suitable parameterization of dynamical variables solves this problem in the ChRMT~\cite{Mollgaard:2014mga}.
They showed that the fermion matrix eigenvalues are well
localized, deviating from the origin,
and the simulation converges to the correct result.
It was also shown in Ref.~\cite{Nagata:2015ijn} that the gauge 
cooling can be extended to the singular drift problem by using suitable norms.

With regard to the convergence problem,
it is interesting to focus on universal quantities to understand 
if the CL method converges correctly.
The CL method can reproduce correct results of the universal 
quantities that are defined as holomorphic quantities 
if the conditions for justification are satisfied~\cite{Aarts:2009uq,Aarts:2011ax}.
On the other hand, it is unclear if the CL method can reproduce 
physical results for non-holomorphic quantities. 
However, it may be natural to expect that the original universality 
is preserved for the correct convergence cases as in one classical thimble calculation
in the LT method \cite{Cristoforetti:2012su}.

The nearest neighbor spacing (NNS) distribution is one of the universal quantities 
which reflect the correlations of the fermion matrix eigenvalues
and the property of the matrix. 
The NNS distribution has been investigated in the phase quenched QCD at zero
and non-zero chemical potential by using MC simulations \cite{Markum:1999yr}.
They showed that the NNS distribution obeys some kinds of distributions:~the
Ginibre, Wigner, and Poisson distributions,
depending on the system and the strength of the eigenvalue correlations.
We note that the NNS distribution is obtained from the density of the fermion matrix eigenvalues
and it is not a holomorphic quantity~\cite{Stephanov:1996RM}.
This fact implies that the distribution obtained in the CL method does not
necessarily reproduce the NNS distribution in the original theory~\cite{Nagata:2015ijn}.
However, the Dirac eigenvalues are closely related to the convergence property of 
the CL method, and the NNS distribution is a universal quantity obtained from Dirac 
eigenvalues. 
Therefore, it can be valuable to investigate the universal behavior of the NNS
distribution of the Dirac eigenvalues in the CL method
deeply to understand the complexified theory.

We conjecture that the NNS distribution can reproduce physical 
results in the CL method even though it is non-holomorphic. 
This conjecture may be natural if the NNS distribution 
reflects the properties of the critical points (or saddle points) owing to its universality. 
Since the Dirac eigenvalues can depend on the choice of the convergence properties
of the CL method~\cite{Mollgaard:2014mga}, then at least we can expect that 
the convergence property of the CL method also affect the NNS distribution. 
In the CL simulations, configurations fluctuate
around a classical flow and locate around some critical points.
The different convergence properties indicate that the configurations
locate around different critical points.
Then, we expect that the underlying universality at each critical point
can be different in the two cases with the correct and wrong convergences
in ChRMT \cite{Mollgaard:2013qra,Mollgaard:2014mga}.

In this work, we investigate the NNS distributions of the ChRMT 
in the CL method, 
where we employ two parameterizations based on
Refs.~\cite{Mollgaard:2013qra,Mollgaard:2014mga}: 
one leads to correct convergence and the other leads to wrong one.
Our purposes are following:
(I) analyzing what happens in the wrong and correct convergent cases in the NNS distribution
and (II) comparing the CL result with the previous MC expectations for
the NNS distribution.

This paper is organized as follows.
In the next section, we define the ChRMT and introduce the two parameterizations. 
In \secref{Sec:Formalism}, we introduce the CL method.
We also explain the procedure for calculating the 
NNS distribution with complex eigenvalues~\cite{Markum:1999yr} in \secref{Sec:NNS}. 
In \secref{Sec:Results}, we give results on the NNS distribution and discussions of them.
\secref{Sec:Summary} is devoted to a summary.

\section{Chiral Random Matrix Theory}
\label{Sec:Formalism}
\subsection{Two types of representations}

In this work, we study the ChRMT introduced in
Ref.~\cite{Osborn:2004rf} in 
two types of the representation: first one is used in Ref.~\cite{Mollgaard:2013qra}, 
and the second one is used in Ref.~\cite{Mollgaard:2014mga}.
Those two representations are equivalent under the linear transformation of dynamical variables
before complexification \cite{Bloch:2012bh}. 
In both cases, we describe originally complex dynamical variables 
in the polar coordinate. 

Throughout this work, we consider only the case of zero topological-index and of 
two flavor $N_f=2$. 
The partition function of the ChRMT is given as
\begin{align}
Z &= \int d\Phi_1 d\Phi_2 (\det (D +m) )^{N_f}
e^{-N \tr (\Phi_1^{\dagger} \Phi_1 + \Phi_2^{\dagger} \Phi_2 )}
\nn \\
&= \int d \Phi_1 d \Phi_2 \exp (-S) \ .
\end{align}
$\Phi_{1,2}$ are $N\times N$ complex matrices.

\vspace{3mm}
\underline{Representation (I): Hyperbolic type}
\vspace{3mm}

First, we consider the following case, where the action and the Dirac operator are
\begin{align}
&S = N \Tr [\Phi_1^{\dagger} \Phi_1 + \Phi_2^{\dagger} \Phi_2]
   - N_f \Tr \log (G^{-1})
\ , \\
&D (\mu)+m = \left(
\begin{matrix}
0 & X \\
Y & 0
\end{matrix}
\right) +m
\ ,
\end{align}
here $G^{-1} = m^2 -XY$.
The $X$ and $Y$ are complex $N \times N$ matrices,
\ba
 X &\equiv i \cosh (\mu) \Phi_1 + \sinh (\mu) \Phi_2^\dagger 
 \ , \\
Y & \equiv i \cosh (\mu) \Phi_1^\dagger + \sinh (\mu) \Phi_2
\ .
\end{align}
We call this parameterization hyperbolic (Hyp) representation below. 
Here, elements of the random matrices, $\Phi_1$ and $\Phi_2$, are originally complex, and
can be parameterized in terms of two real variables.
In the present study, we employ the polar coordinate~\cite{Mollgaard:2014mga}, where
\ba
\PhiM{1}{i j} = \rM{1}{i j} e^{i \tM{1}{i j} } 
 \ \ ,\ 
\PhiM{2}{i j} = \rM{2}{i j} e^{i \tM{2}{i j} } 
\ ,
\end{align}
for $i,j = 1,\cdots, N$. The matrices $X$ and $Y$ are given as
\ba
X_{ij} &= i \cosh (\mu) \rM{1}{i j} e^{ i \tM{1}{ij} }
        + \sinh (\mu) \rM{2}{j i} e^{-i \tM{2}{j i}}
\ , \\
Y_{i j} & = i \cosh (\mu) \rM{1}{j i} e^{- i \tM{1}{ji}}
          + \sinh (\mu) \rM{2}{i j} e^{i \tM{2}{i j}}
\ .
\end{align}
The action is rewritten as
\begin{align}
 S =& - \sum_{i,j} \left( \log (\rM{1}{i j}) + \log (\rM{2}{i j}) \right) 
   \nn \\
  & - \Nf \log \det (G^{-1}) 
  + N \sum_{i,j} \left(  \rM{1}{ij}^2 + \rM{2}{ji}^2 \right)
  \ .
\label{Eq:action_HYP}
\end{align}

\underline{Representation (II): Exponential type}
\vspace{3mm}

Next, we consider the other parameterization used in Ref.~\cite{Mollgaard:2014mga} as
\begin{align}
X \equiv & e^\mu \Phi_1 - e^{-\mu} \Phi_2^\dagger \ , \\
Y \equiv & -e^{-\mu} \Phi_1^\dagger + e^\mu \Phi_2\ .
\end{align}
The action is given by 
\begin{align}
S = 2 N {\rm Tr} [ \Phi_1^\dagger \Phi_1 + \Phi_2^\dagger \Phi_2] - N_f {\rm Tr} \log (G^{-1}).
\end{align}
This parameterization is referred to as Exponential (Exp) representation below.
We also use the polar coordinate in this representation. 

The two representations are equivalent under a linear transformation \cite{Bloch:2012bh,Mollgaard:2014mga}.
Note that the CL method is applied to the ChRMT with the Hyp representation and with the Cartesian 
coordinate in Ref.~\cite{Mollgaard:2013qra}, while it is applied to the ChRMT with 
the Exp representation and with the polar coordinate in Ref.~\cite{Mollgaard:2014mga}. 
In this work, we adopt the polar coordinate both for the two representations.
We will confirm that only the case of the Exp representation reproduces
the correct results if the mass of quark is small at non-zero quark chemical potential.

\subsection{Langevin equations and drift terms}

Now, we apply the Langevin equation to the originally real variables 
in the system $\varphi_k \in
\{ r_{1, ij} , r_{2, ij} , \theta_{1,ij}, \theta_{2,ij}|\,   i, j = 1, \cdots, N, ( i, j \in \mathbb{Z})\}$.
\begin{align}
\frac{\partial \varphi_k (\tau)}{\partial \tau} = - \frac{\partial S [\varphi]}{\partial \varphi_k} + \eta_k (\tau)
\ ,
\end{align}
where $\tau$ is the Langevin time. 
The Gaussian noise, $\eta_k$, is normalized as $\expv{\eta_k (\tau )} = 0$
and $\expv{\eta_k (\tau) \eta_l (\tau')} = 2 \delta_{kl} \delta (\tau -\tau') $.
Since the action is complex at $\mu \neq 0$, those variables are extended to complex
$r_1, r_2, \theta_1, \theta_2 \in \mathbb{R} \to \mathbb{C}$. 
In this paper, we use the real Gaussian noise $\eta_k \in \mathbb{R}$
and the adaptive step size method
\cite{Ambjorn:1986fz,Aarts:2009dg,Aarts:2010aq}. 
We explicitly show the drift terms in the Hyp 
representation in \secref{sec:driftterms}, and those in the Exp representation
were shown in Ref.~\cite{Mollgaard:2014mga}.

In the numerical simulation, we define the discretized Langevin equation
\begin{align}
\varphi_k (\tau+d\tau) =  \varphi_k (\tau) + d \tau \biggl(- \frac{\partial S [\varphi]}{\partial \varphi_k} 
\biggr) + \sqrt{d\tau}\, \eta_k (\tau).
\end{align}

In the adaptive stepsize method in Ref.~\cite{Aarts:2010aq},
we can adopt an average value of a maximum drift term 
at each Langevin time before the thermalization time $\tau_{\rm th}$,
\begin{align}
\left< \mathcal{K}_\mrm{max} \right>_\mrm{th}= 
\frac{1}{N_\mrm{th}}  \sum_{\tau' =0}^{\tau_\mrm{th}}
\mathcal{K}_\mrm{max} (\tau')
\ ,
\end{align}
where
$\mathcal{K}_\mrm{max} (\tau') 
= \max_{k,\varphi=r_1,r_2,\theta_1,\theta_2} \left| \varphi_k (\tau') \right| $
and 
$N_\mrm{th} = \tau_\mrm{th}/d\tau$ is the number of configurations
until the thermalization time, $\tau_\mrm{th}$.
After the thermalization time,
we adopt an adaptive stepsize at each Langevin time as
\begin{align}
\epsilon_{\tau} = \min (d\tau,  d\tau \left< \mathcal{K}_\mrm{max} \right>_\mathrm{th} / \mathcal{K}_\mrm{max} (\tau) )
\ , 
\end{align}
and then the CL equation is given as
\begin{align}
\varphi_k(\tau + \epsilon_\tau) 
= \varphi_k (\tau) + \epsilon_\tau \left(- \frac{\partial S [\varphi]}{\partial \varphi_k} \right)+ \sqrt{\epsilon_\tau} \eta_k (\tau) \ .
\end{align}



\section{Nearest Neighbor spacing distribution} \label{Sec:NNS}

In order to obtain the nearest neighbor spacing (NNS) distributions,
we need a procedure so called unfolding.
We follow the unfolding procedure introduced in Ref.~\cite{Markum:1999yr}.

\subsubsection{Unfolding procedure with complex eigenvalues}

First, we define the density of Dirac eigenvalues as
\begin{align}
 \rho (x,y) = \expv{ \sum_k \delta^{(2)} ( z - z_k )} \ , \, ( z = x + i y)
\end{align}
where $\expv{\cdots}$ denotes an ensemble average 
and $z_k$ is an eigenvalue of the Dirac operator
$ D(\mu) \ket{z_k} =  z_k \ket{z_k},\, (k=1, \cdots 2N)$.
Due to the property of the Dirac operator $\{ \gamma_5, D(\mu)\}=0$,
the eigenvalues appear as $\pm z$ pair \cite{Akemann:2004dr}.
The density of the Dirac eigenvalue $\rho(x,y)$ is non-holomorphic~\cite{Stephanov:1996RM}.

In order to obtain the fluctuation part of the 
eigenvalue density, we rewrite $\rho(x,y)$ as 
\begin{align}
\rho(x,y) = \rho_{\mrm{ave} } (x,y)+ \rho_{\mrm{fluc}}(x,y) \ ,
\end{align}
where $\rho_{\mrm{ave} }$ and $\rho_{\mrm{fluc}}$ are the average part and the 
fluctuation part of the eigenvalue density, respectively.
We consider a map 
\begin{align}
z' = x' + i y' = u(x,y) +i v(x,y)
\label{eq:2016feb16eq1}
\ , 
\end{align}
where we impose a condition that the average eigenvalue density is unity for any points on the new coordinate, 
namely $\rho_{\mrm{ave} } (x',y')=1$.
It immediately follows from the probability conservation condition that
$\rho_{\mrm{ave} } (x',y')dx'dy'=dx'dy' =\rho_{\mrm{ave}} (x,y)dxdy$ \cite{Markum:1999yr}.
On the other hand, it follows from Eq.~(\ref{eq:2016feb16eq1})
that $dx' dy' = J dx dy,$ where $J=|\partial (x',y')/\partial(x,y)|$ is the Jacobian 
of the coordinate transformation. Combining the two relations implies that the 
$\rho_{\mrm{ave}} (x,y)$ is nothing but the Jacobian, i.e., 
\begin{align}
\rho_{\mrm{ave}} (x,y) = \left|  \partial_x u \partial_y v - \partial_x v \partial_y u  \right|   
\ .
\label{eq:2016feb16eq2}
\end{align}
$u(x,y)$ and $v(x,y)$ are not determined uniquely only from this condition.
We choose $y'=v(x,y)=y$, which reduces Eq.~(\ref{eq:2016feb16eq2}) to 
$\rho_{\mrm{ave}} (x,y) = \partial_x u $ \cite{Markum:1999yr}.
The real part of eigenvalues is expressed as
\begin{align}
  x' = u(x,y) = \int_{-\infty}^{x} dt \rho_{\mrm{ave}} (t,y) \equiv N_{\mrm{ave}} (x,y)
  \ ,
\end{align}
where $N_{\mrm{ave}}(x,y)$ is the average part of the cumulative
spectral function $N(x,y)$~\cite{Guhr:1997ve}, which is given as
\begin{align} 
 N (x,y ) = N_{\mrm{ave}} (x,y) + N_{\mrm{fluc}} (x,y)
 \ .
\end{align}

\begin{figure}[htbp]
 \includegraphics[width=60mm, angle=270]{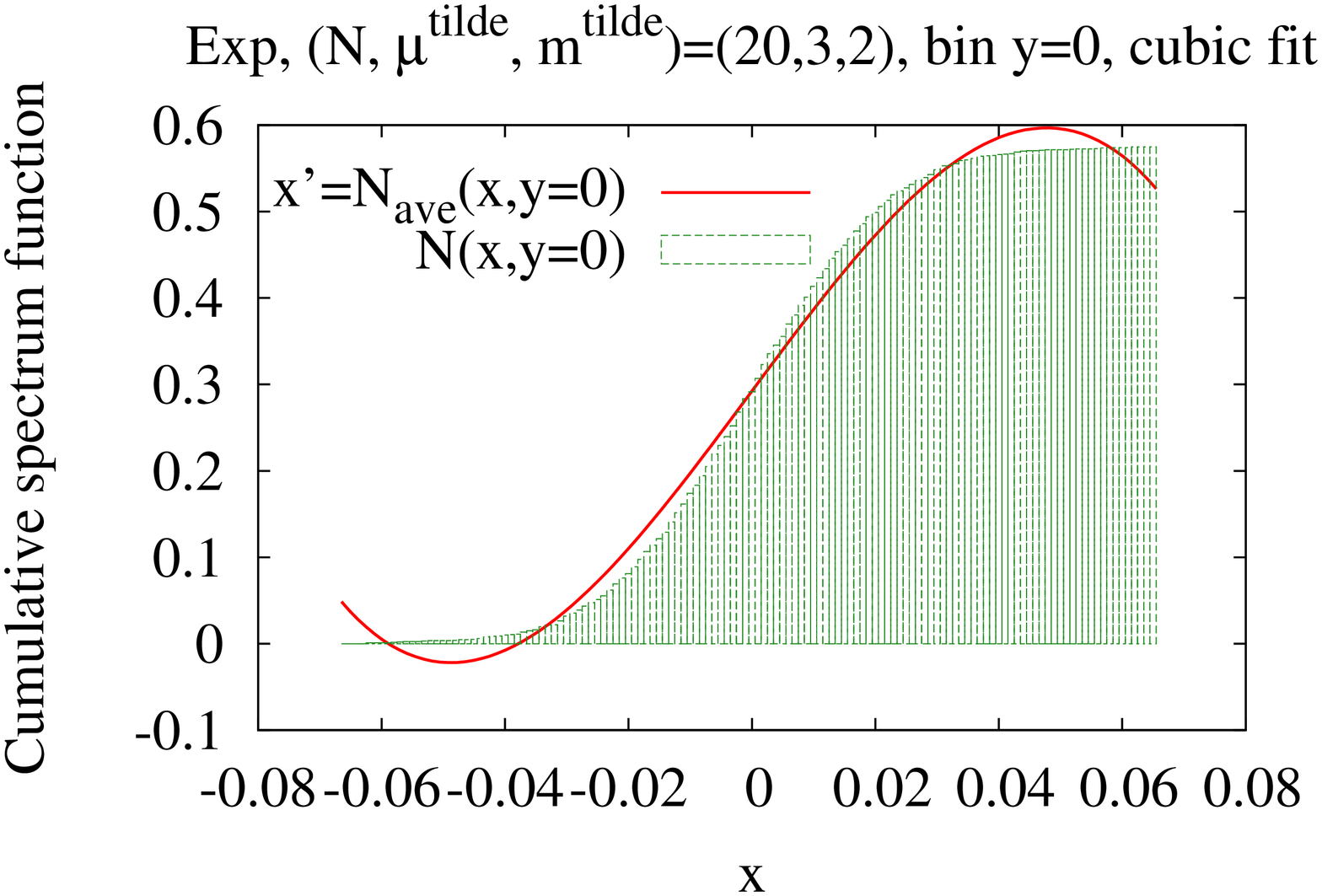}
   \includegraphics[width=60mm, angle=270]{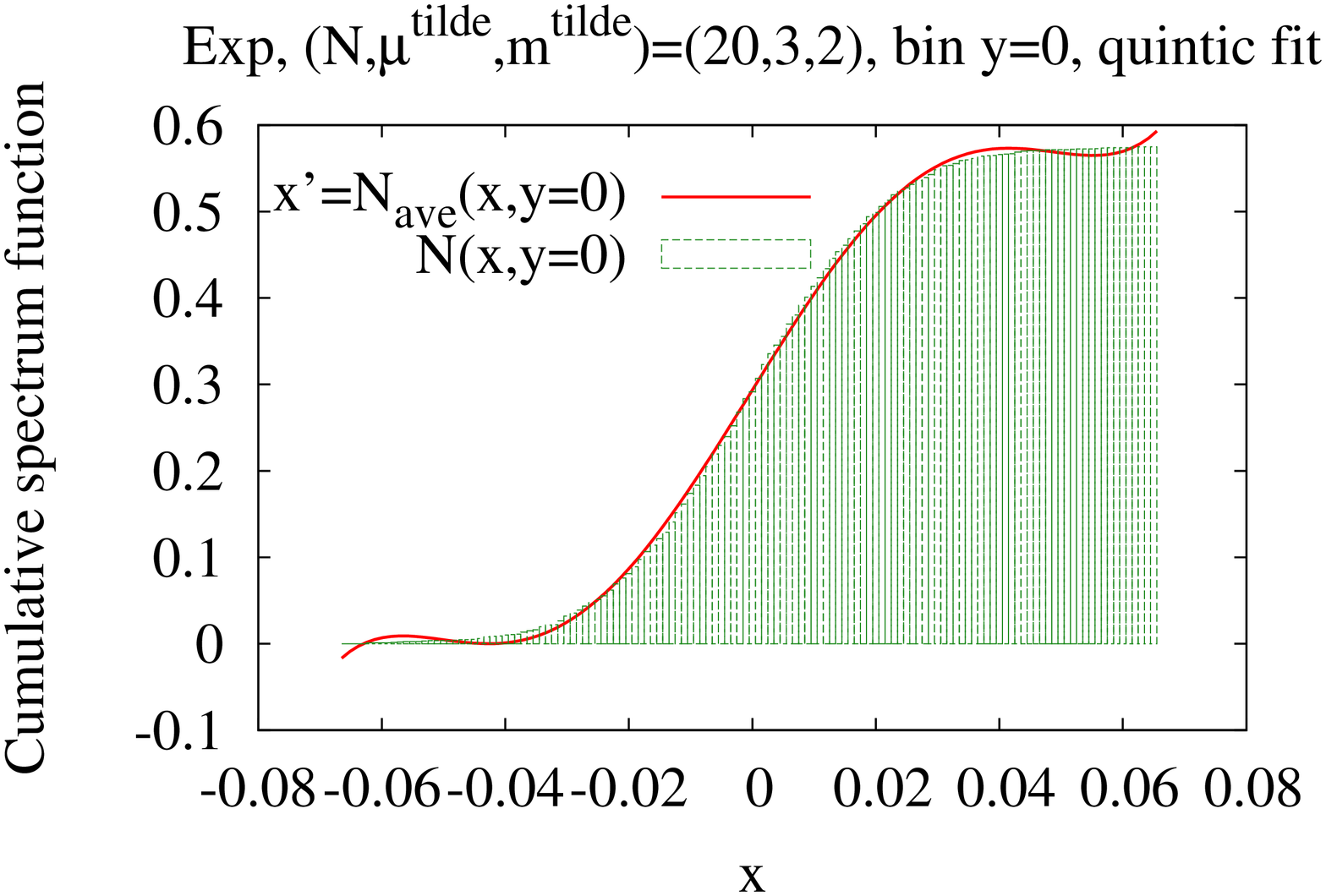}
  \caption{The cumulative spectral function $N(x,y)$ (histogram) and its 
average part $N_{\rm ave}(x,y)$ (solid line) as a function of $x$ 
at $N=20$, $\tilde{m}=2$, and $\tilde{\mu}=3$ in the Exp representation. 
They are obtained from eigenvalues included in a strip around $y=0$. 
The top and bottom panels are obtained with a cubic function 
and a quintic function, respectively.
 \label{fig:Cumu-N20mu3m2}
  }  \label{fig:Unfolding-N20mu3m2}
\end{figure} 

In order to obtain the cumulative spectral function numerically,
we first divide the complex plane in the $y$-direction, where 
each strip has a width $dy$. 
In the ChRMT, a spacing between two adjacent eigenvalues 
is of an order $\mathcal{O}(1/N)$ \cite{Verbaarschot:2000dy}. 
We choose the width $dy$ so that it is bigger than this magnitude:
$dy = 0.1$ at $N=20$, for example.
If we adopt too small $dy$, there are few eigenvalues in a given strip.
Then, we calculate $N(x,y)$ for each strip.
In our calculations, we derive the average cumulative spectral function
by fitting $N_{\mrm{ave}}$ with a low order polynomial as in Ref.~\cite{Markum:1999yr}.
Then, we adopt a sufficiently small bin size for $x$
to fit $N_\mrm{ave}$ appropriately.
We use a quintic polynomial throughout this paper, but the
validity of the fitting procedure is checked by comparing results with
the quintic and cubic polynomials.
In \figref{fig:Cumu-N20mu3m2}, we show $N_{\rm ave}(x,y)$ obtained from
the cubic polynomial (top panel), and from the
quintic polynomial (bottom panel).
We find that the NNS distribution is quantitatively insensitive to the choice 
of the fitting functions, which will be shown in Appendix \ref{Sec:Fit}.



\begin{figure}[htbp]
   \includegraphics[width=30mm, angle=270, bb= 0 0 595 842]{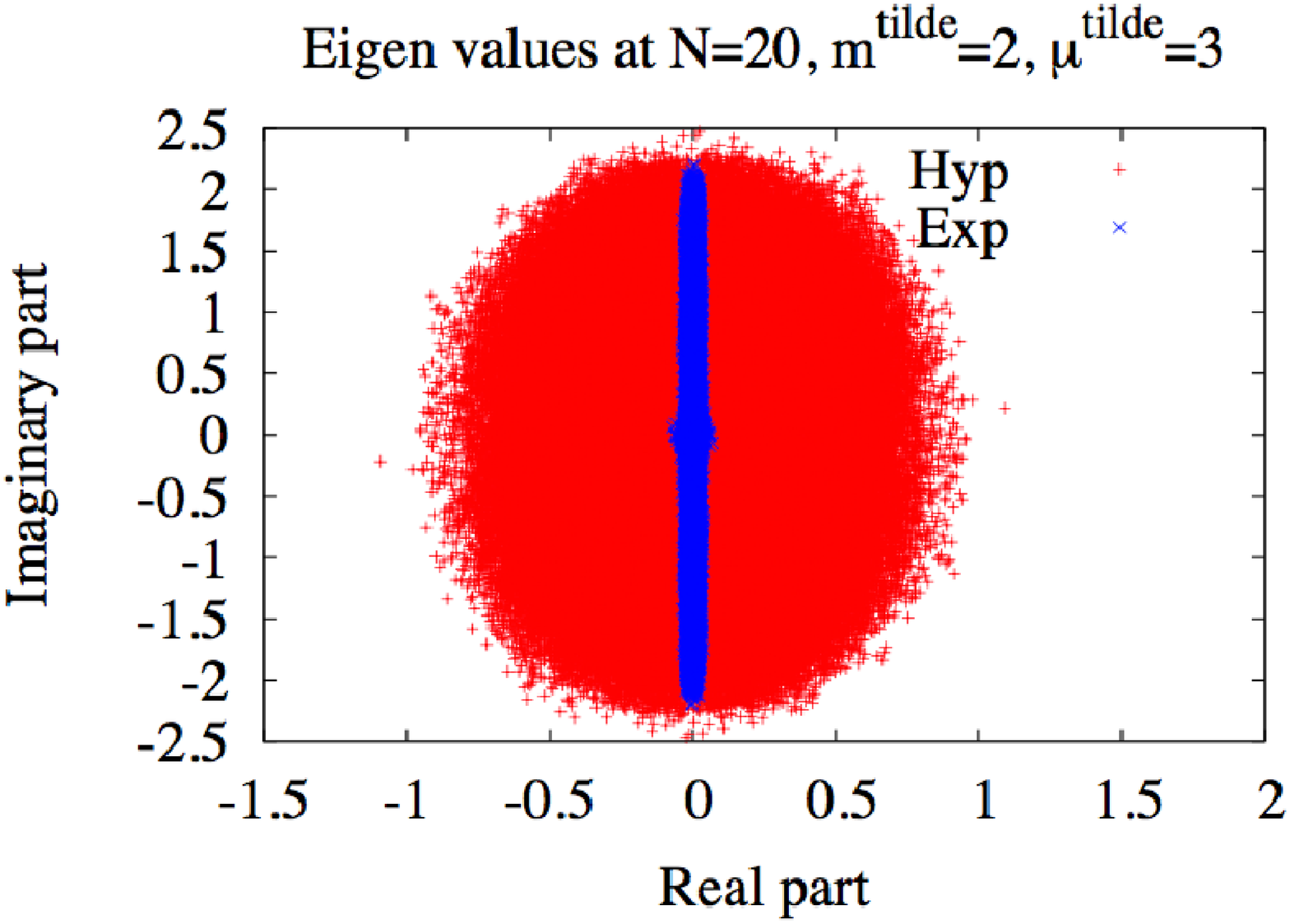}
   \includegraphics[width=30mm, angle=270, bb= 0 0 595 842]{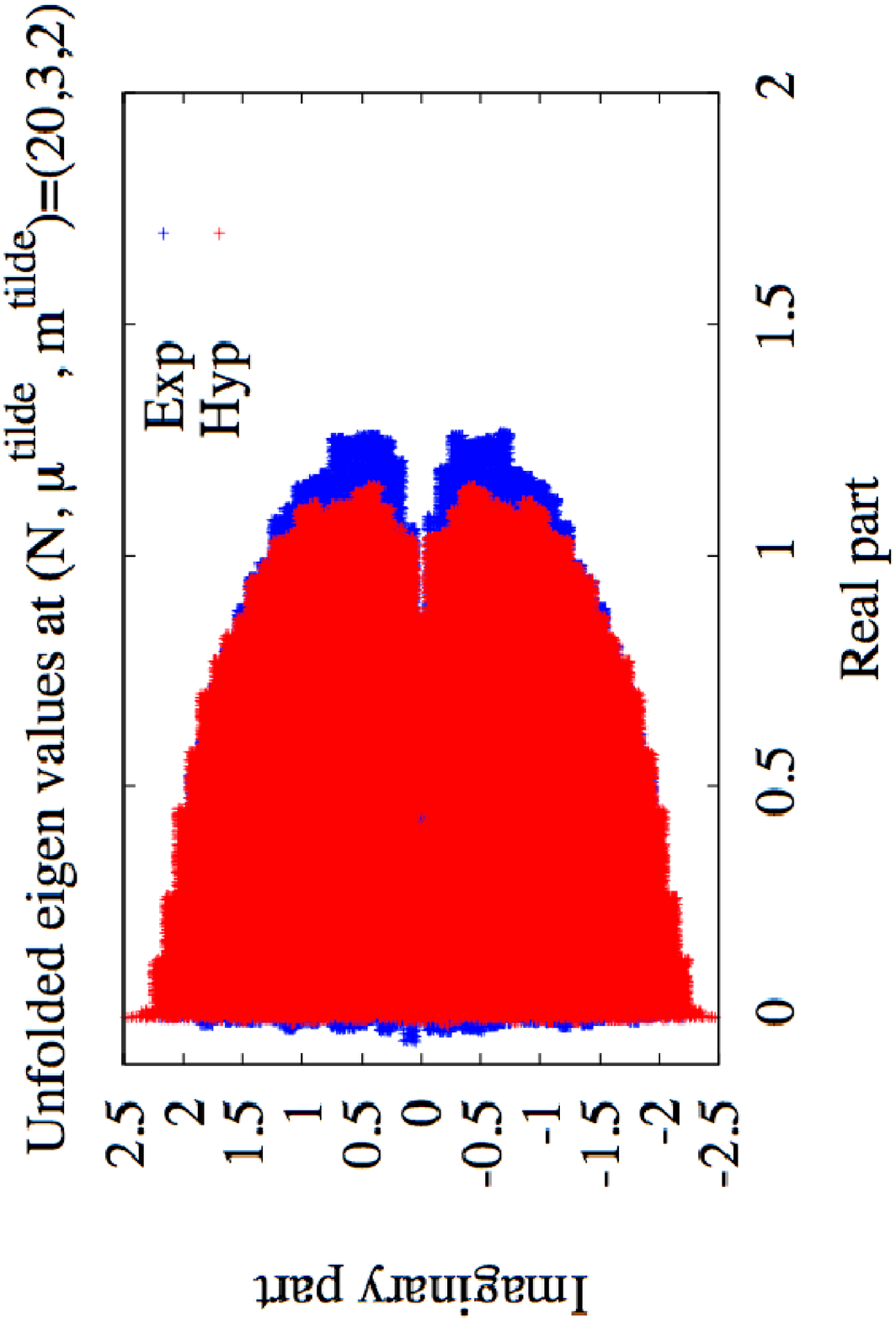}
  \caption{
  Dirac eigenvalues
  at $N=20$, $\tilde{m}=2$, and $\tilde{\mu}=3$
  before unfolding procedure (left panel) and after unfolding procedure (right panel).
  They are obtained from configurations in late Langevin time.
  \label{fig:EV-N20mu3m2}
  }
\end{figure}


To illustrate the unfolding procedure, we show Dirac eigenvalues at $N=20$,
$\tilde{m}=2$, and $\tilde{\mu}=3$
obtained with various ensembles at late Langevin times  in \figref{fig:EV-N20mu3m2}
where $\tilde{m}=Nm$ and $\tilde{\mu}=\sqrt{N}\mu$.
As in Ref.~\cite{Mollgaard:2014mga},
the width of Dirac eigenvalues in the Exp representation
decreases along the real axis compared with that in the Hyp representation.
The right panel shows the Dirac eigenvalues after the unfolding.
The unfolded distribution is apparently consistent with the
distribution obtained in the phase quenched QCD~\cite{Markum:1999yr}.

\subsubsection{Nearest Neighbor spacing distribution with complex eigenvalues}
In order to derive the NNS distribution from complex eigenvalues,
we need to introduce the NNS, $s$.
There is an ambiguity to define $s$ for the complex
eigenvalues.
We adopt $s_i (\tau) = \min_j \left| z_i' - z_j' \right|$ for
$i = 1, \cdots, 2N$ using unfolded 
eigenvalues at Langevin time $\tau$ \cite{Markum:1999yr}.
We construct the distribution $p(s)$ of $s_i (\tau)$ for $(i =1,\cdots,2N)$,
where we use all the configurations 
in late Langevin time
with a certain interval.

The NNS distribution $P(s)$ is defined so that it satisfies
two conditions $\int ds P(s) = 1$ and $\int ds s P(s) = 1$.
In order to satisfy these conditions, we use the normalization
and rescaling of $s$ \cite{Markum:1999yr}:
supposing that the first moment of the distribution is $c =
\int_0^{\infty} ds s p(s)$ with $1 = \int_0^{\infty} ds p(s)$,
the new distribution is defined by $P(s) = c p(c s)$.
Once the distribution satisfies the two conditions,
we can compare the distribution obtained in the present work
with some typical distributions.

\subsubsection{Reference NNS distributions}
Here, we explain three typical NNS distributions, which are used as references 
to understand numerical results obtained from CL simulations. 

At $\mu=0$, the Dirac operator is anti-Hermitian, and its eigenvalues are pure imaginary. 
The NNS distribution in the chiral unitary ensemble of RMT follows the Wigner surmise~\cite{Halasz:1995vd,Guhr:1997ve,Markum:1999yr}
\begin{align}
P_W(s) = \frac{32}{\pi^2}s^2 e^{-4s^2/\pi}.
\end{align}
It is expected that the NNS distribution in this work also follows the Wigner surmise.
We will discuss the mass dependence in \secref{Sec:Results}.

On the other hand, at $\mu\neq 0$, the Dirac operator is no more anti-Hermitian, 
and its eigenvalues are generally complex. 
If the real and imaginary part of eigenvalues have approximately the same average magnitude, then 
the system is described by the Ginibre ensemble \cite{Ginibre:1965zz} of non-Hermitian RMT~\cite{Ginibre:1965zz,Markum:1999yr}.
In this case, the NNS distribution is given by 
\begin{align}
P_G(s) & = c p(cs) , \\
  p(s) & = 2s \lim_{N \to \infty} \left[ \prod_{n=1}^{N-1} e_n(s^2) e^{-s^2} \right] \sum_{n=1}^{N-1} \frac{s^{2n}}{n! e_n(s^2)}
  \label{eq:ps-anal}, 
\end{align}
where $e_n(x) = \sum_{m=0}^n x^m/m!$ and $c=\int_0^\infty ds s p(s) $ \cite{Grobe:1988zz,Markum:1999yr}. 
In this paper, we use $N=2000$ in \equref{eq:ps-anal} as a reference distribution.

For uncorrelated eigenvalues, the NNS distribution follows the Poisson distribution. 
On the complex plane, the Poisson distribution is given by \cite{Grobe:1988zz}
\begin{align}
P_{P} (s)= \frac{\pi}{2} s e^{-\pi s^2/4}.
\end{align}

\section{Results and discussion}
\label{Sec:Results}

In this section, we show numerical results obtained from the CL simulations. 
Our numerical set up is as follows.
We consider the ChRMT with zero topological index $\nu =0$, $N =20$ and $\Nf =2$.
We utilize the reference step size as $d\tau=1.0\times 10^{-6}$, and 
perform the simulation with the adaptive stepsize~\cite{Aarts:2010aq}. 
We take the Langevin time $\tau \sim 50$ for thermalization, and $\tau \sim 100$ for measurements. 
The measurement is performed by using configurations in late Langevin time, and 
the number of typical configurations for the measurement is 2000.

In the following, we show the numerical results for $\tilde{\mu}=1$ and $\tilde{\mu}=3$, 
where $\tilde{m}=Nm$ and $\tilde{\mu}=\sqrt{N}\mu$.

\subsection{Results for $N=20$ at $\tilde{\mu} =1$}
\begin{figure}[htbp]
 \begin{center}
  \includegraphics[width=60mm, angle=270]{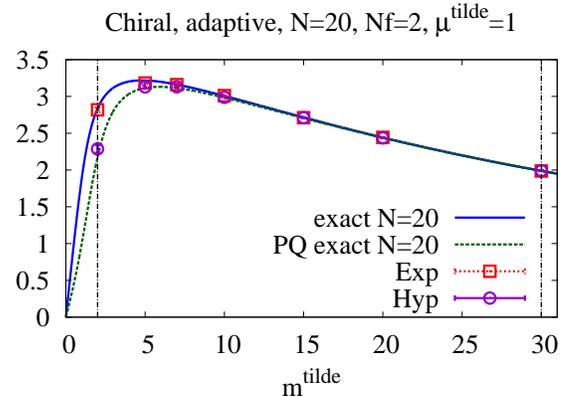}
 \end{center}
 \caption{The chiral condensate at 
 $\tilde{\mu}=1$ as a function of $\tilde{m}$.
 The solid line shows the exact value in the ChRMT \cite{Osborn:2004rf} and
 the dotted line shows the exact value in the phase quenched (PQ)
 ChRMT \cite{Akemann:2004dr}.
 The results of the Exp and Hyp representations are expressed as squares and circles.
 At large $\tilde{m}$, both representations well produce exact results.
 Results of the Hyp representation slightly deviate from exact results at small $\tilde{m}$.
 Dotted vertical lines show the values of $\tilde{m}$ at which the NNS distribution 
 is obtained.}
 \label{fig:N20-chiral-mu1}
\end{figure}

In \figref{fig:N20-chiral-mu1}, we show the chiral condensate at $\tilde{\mu}=1$ as a function of $\tilde{m}$. 
Here the chiral condensate is given by \cite{Mollgaard:2013qra}, 
\begin{align}
\frac{1}{N} \expv{\bar{\eta} \eta} &= \frac{1}{N} \expv{ \partial_m \log (Z)  }
\nn \\
& = \frac{2m \Nf}{N} \expv{ \Tr \left[ (m^2 - XY)^{-1} \right] }
\ .
\label{Eq:chiral}
\end{align}

There is a small difference between analytical results for the exact \cite{Osborn:2004rf} and phase quenched \cite{Akemann:2004dr} cases for small $\tilde{m}$, which are denoted as ``exact" and ``PQ exact" 
in \figref{fig:N20-chiral-mu1}, respectively.
Numerical results with the Exp representation and Hyp representation are 
almost consistent with analytical results for the exact and phase quenched cases.


\begin{figure}[htb]
  \begin{center}
   \includegraphics[width=60mm, angle=270]{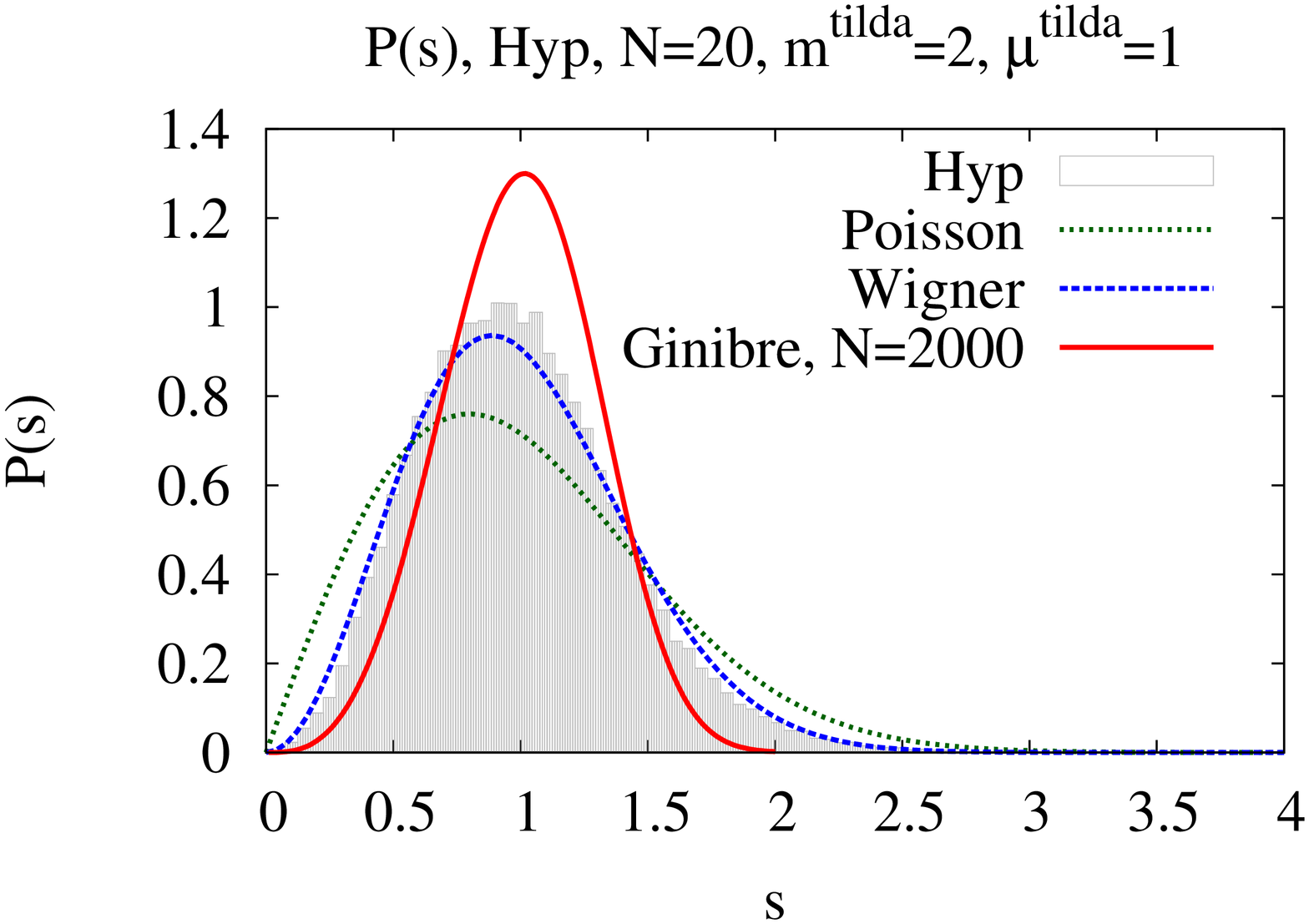}
  \end{center}
  \caption{
  The NNS distribution of
  the Hyp notation at $N=20$, $\tilde{m}=2$, and $\tilde{\mu}=1$.}
  \label{fig:N20mu1m2H}
  \begin{center}
   \includegraphics[width=60mm, angle=270]{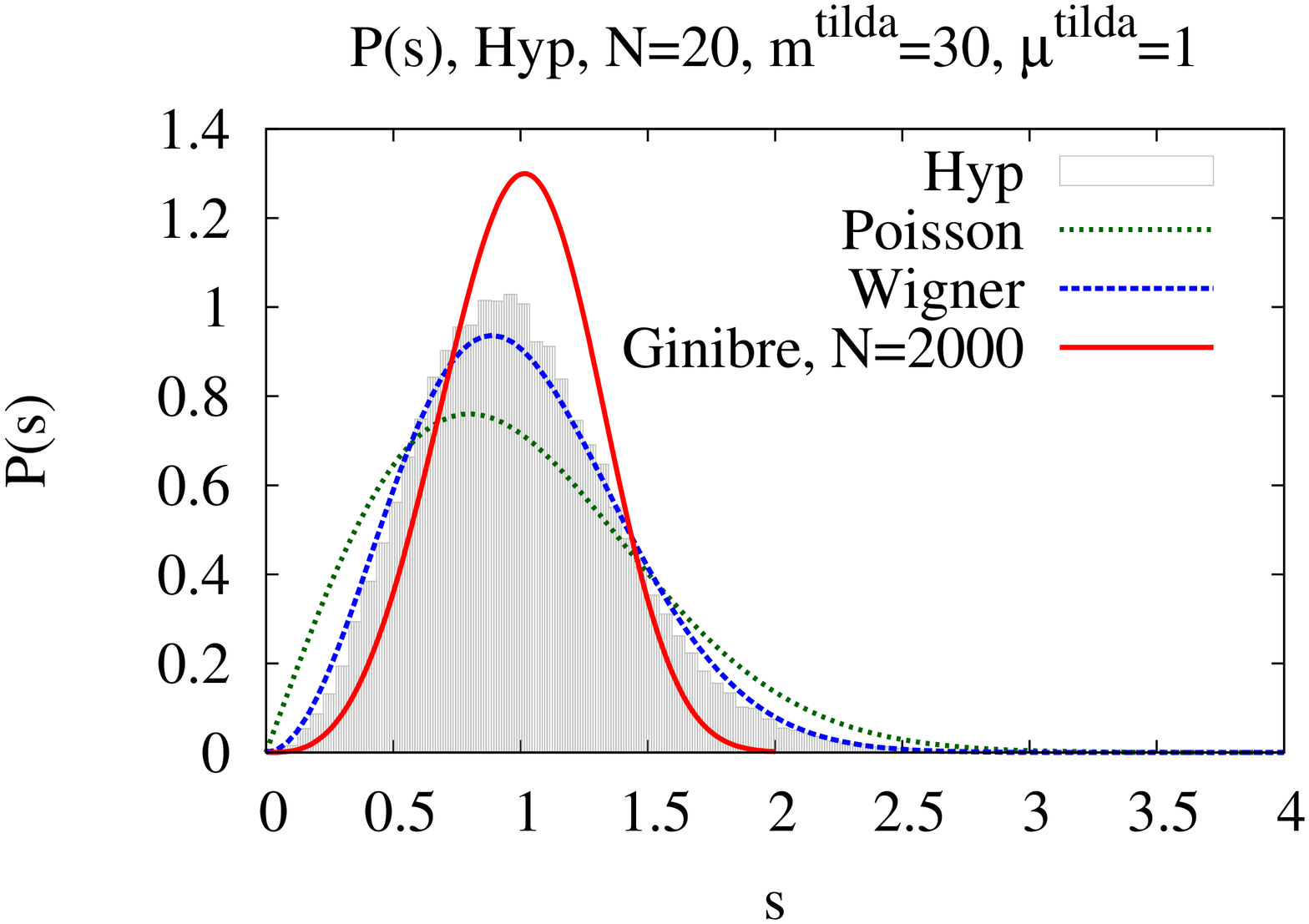}
  \end{center}
  \caption{
  The NNS distribution of 
  the Hyp notation at $N=20$, $\tilde{m}=30$, and 
  $\tilde{\mu}=1$.}
  \label{fig:N20mu1m30H}
\end{figure}

We show the NNS distribution in the Hyp representation
for ($\tilde{\mu}, \tilde{m})=(1,2)$ and (1,30) in
Figs.~\ref{fig:N20mu1m2H} and \ref{fig:N20mu1m30H}, respectively.
The NNS distributions at $\tilde{m}=2$ and $30$ are almost
consistent with the Wigner surmise.


\begin{figure}[htb]
  \begin{center}
   \includegraphics[width=60mm, angle=270]{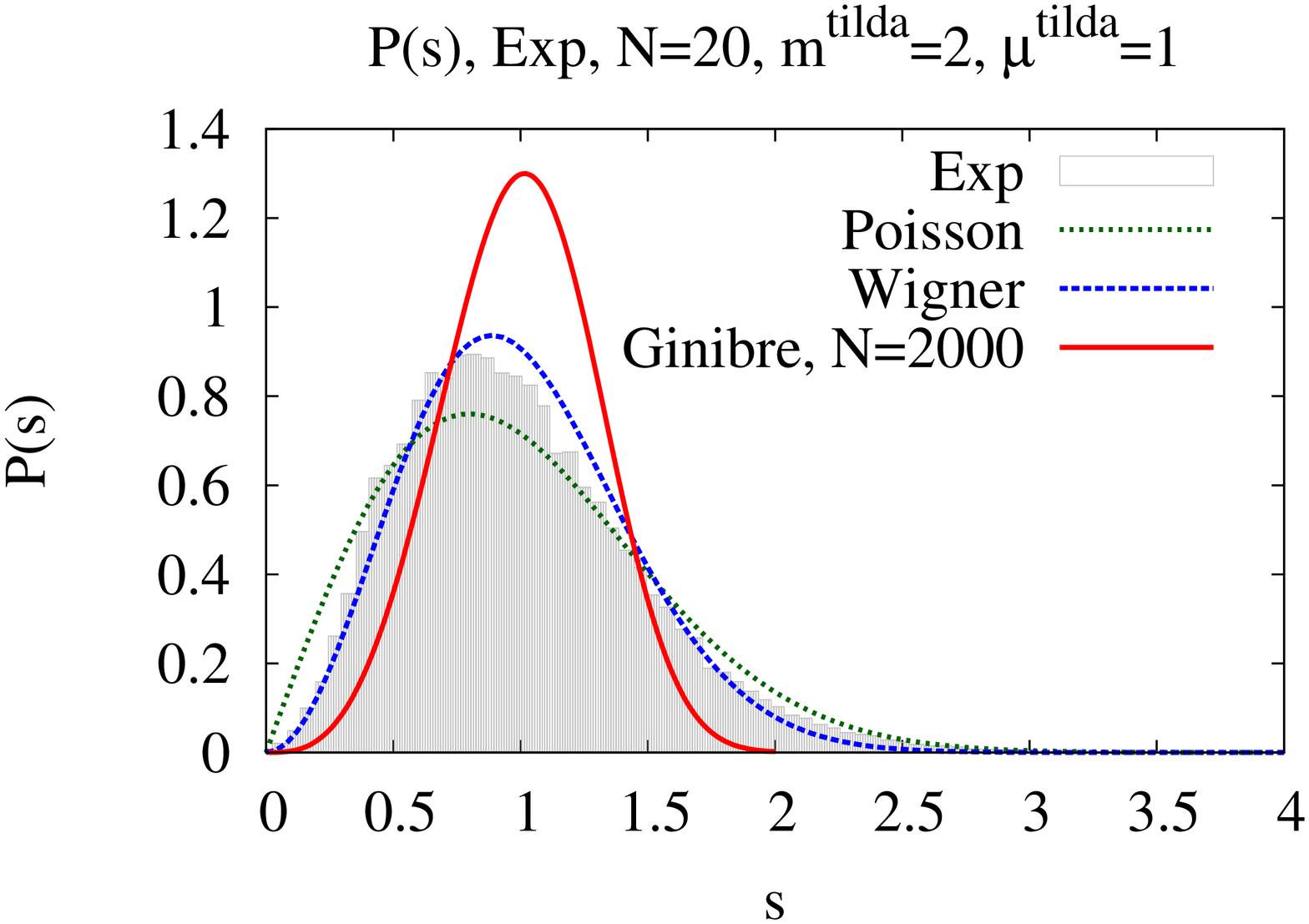}
  \end{center}
  \caption{
  The NNS 
  distribution of 
  the Exp notation at $N=20$, $\tilde{m}=2$, and $\tilde{\mu}=1$.}
  \label{fig:N20mu1m2E}
  \begin{center}
   \includegraphics[width=60mm, angle=270]{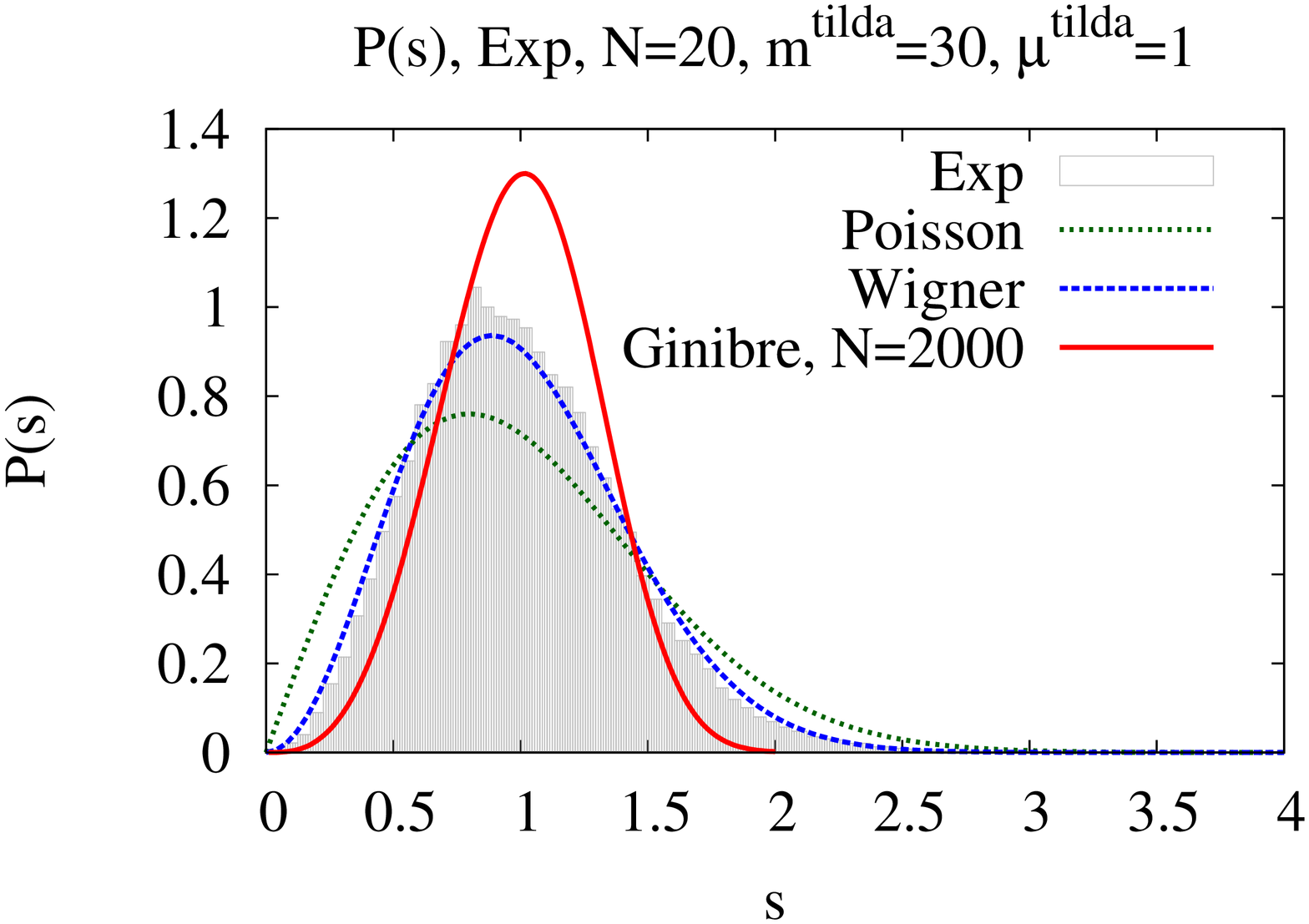}
  \end{center}
  \caption{
  The NNS 
  distribution of 
  the Exp notation at $N=20$, $\tilde{m}=30$, and $\tilde{\mu}=1$.}
  \label{fig:N20mu1m30E}
\end{figure}


We show the NNS distribution in the Exp representation
for ($\tilde{\mu}, \tilde{m})=(1,2)$ and (1,30) in
Figs.~\ref{fig:N20mu1m2E} and \ref{fig:N20mu1m30E}, respectively.
For $\tilde{m}=30$, the NNS distribution in the Exp representation is almost
the same as that in the Hyp representation.
By comparison, the NNS distribution at $\tilde{m}=2$ is slightly different from
NNS distribution at $\tilde{m}=30$ and is relatively close to the Wigner surmise.


%

For small $\tilde{m}$, the NNS distributions in the two representations
are slightly different.
The difference increases for $\tilde{\mu}=3$ as we will
show in the next subsection.

\subsection{Results for $N=20$ at $\tilde{\mu} =3$}
\label{Sec:results}

\begin{figure}[htbp]
 \begin{center}
  \includegraphics[width=60mm, angle=270]{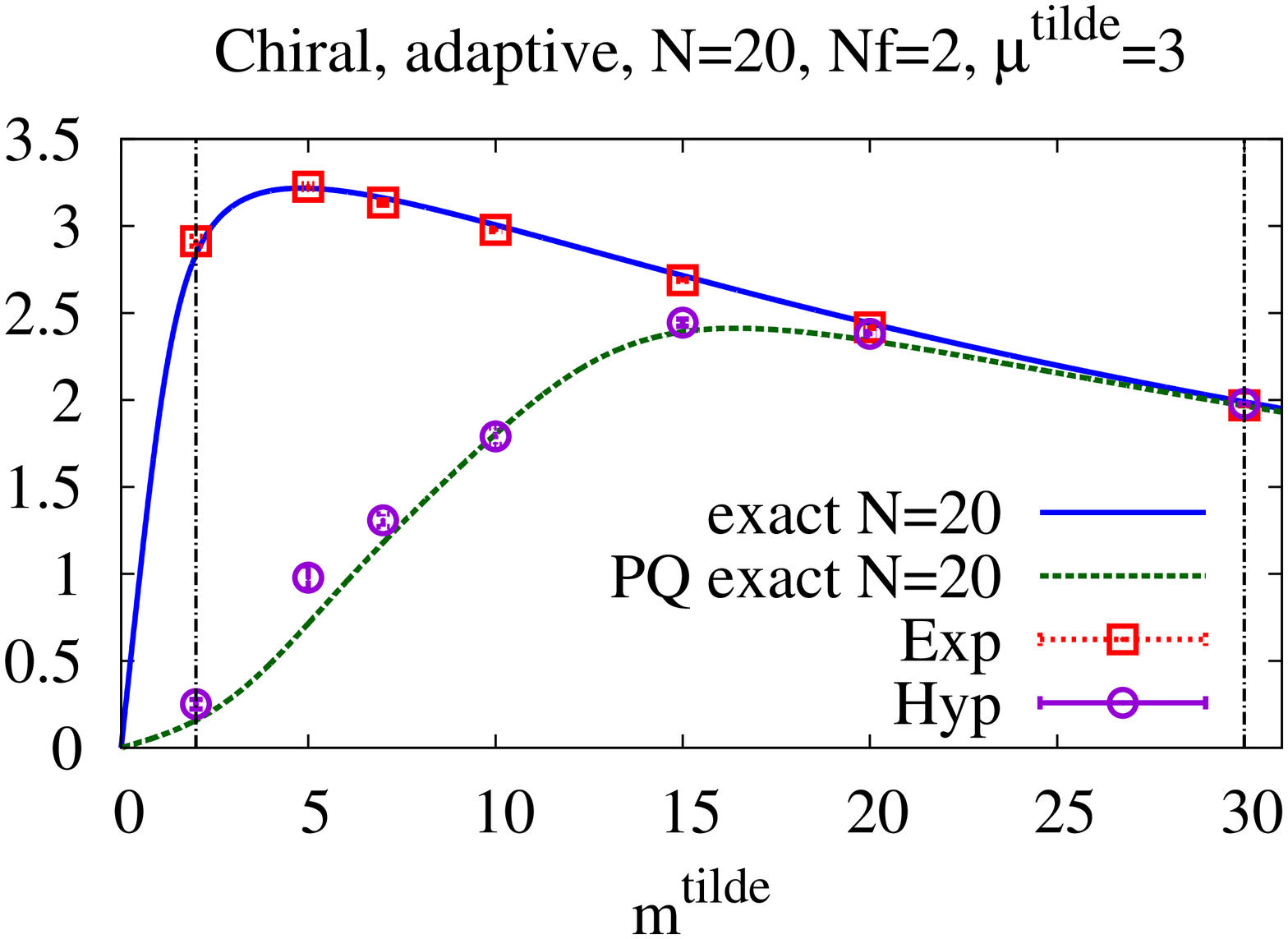}
 \end{center}
 \caption{Chiral condensate at $\tilde{\mu}=3$ as a function of $\tilde{m}$.
 The solid line shows the exact value in the ChRMT \cite{Osborn:2004rf} and
 the dotted line shows the exact value in the phase quenched (PQ)
 ChRMT \cite{Akemann:2004dr}.
 The results of the Exp and Hyp representations are expressed as
 squares and circles.  At large $\tilde{m}$, both the representations well produce the exact result.
 The result of Hyp representation at small $\tilde{m}$ deviates from the exact result.
 Dotted vertical lines show the values of $\tilde{m}$ at which the NNS distribution 
 is obtained.
}
 \label{fig:N20-chiral}
\end{figure}

In \figref{fig:N20-chiral},
we show the chiral condensate at $\tilde{\mu} = 3$ as a function of $\tilde{m}$.
The chiral condensate in the Exp representation reproduces the correct result
shown by the solid line, as found in \cite{Mollgaard:2014mga}.
On the other hand, the chiral condensate in the Hyp representation
produces results close to the phase quenched theory 
 shown by dotted line at low $\tilde{m}$.


\begin{figure}[htbp]
  \begin{center}
   \includegraphics[width=60mm, angle=270]{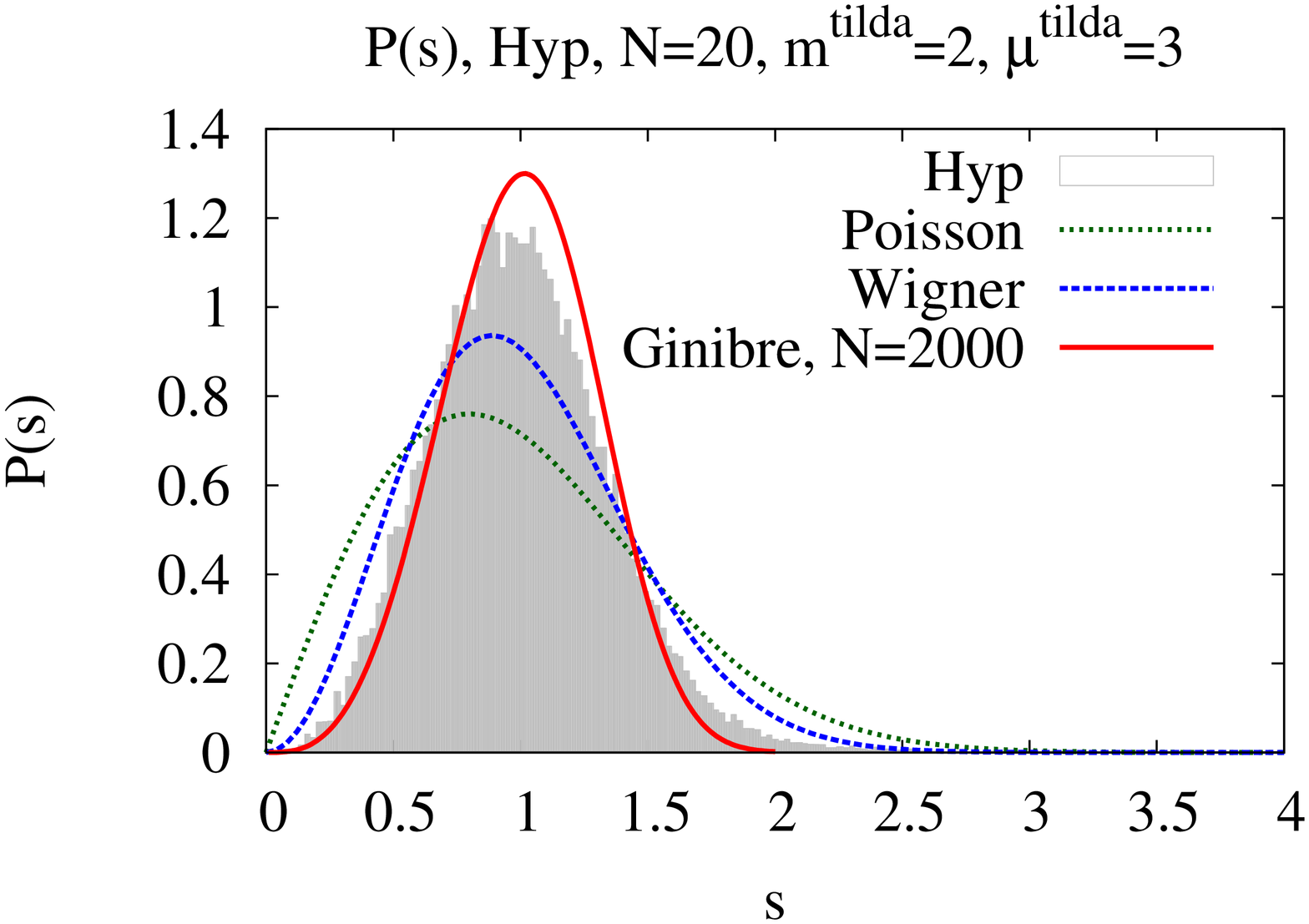}
  \end{center}
  \caption{
  The NNS distribution of 
  the Hyp notation
  at $N=20$, $\tilde{m}=2$, and $\tilde{\mu}=3$.
  }
  \label{fig:N20mu3m2H}
  \begin{center}
   \includegraphics[width=60mm, angle=270]{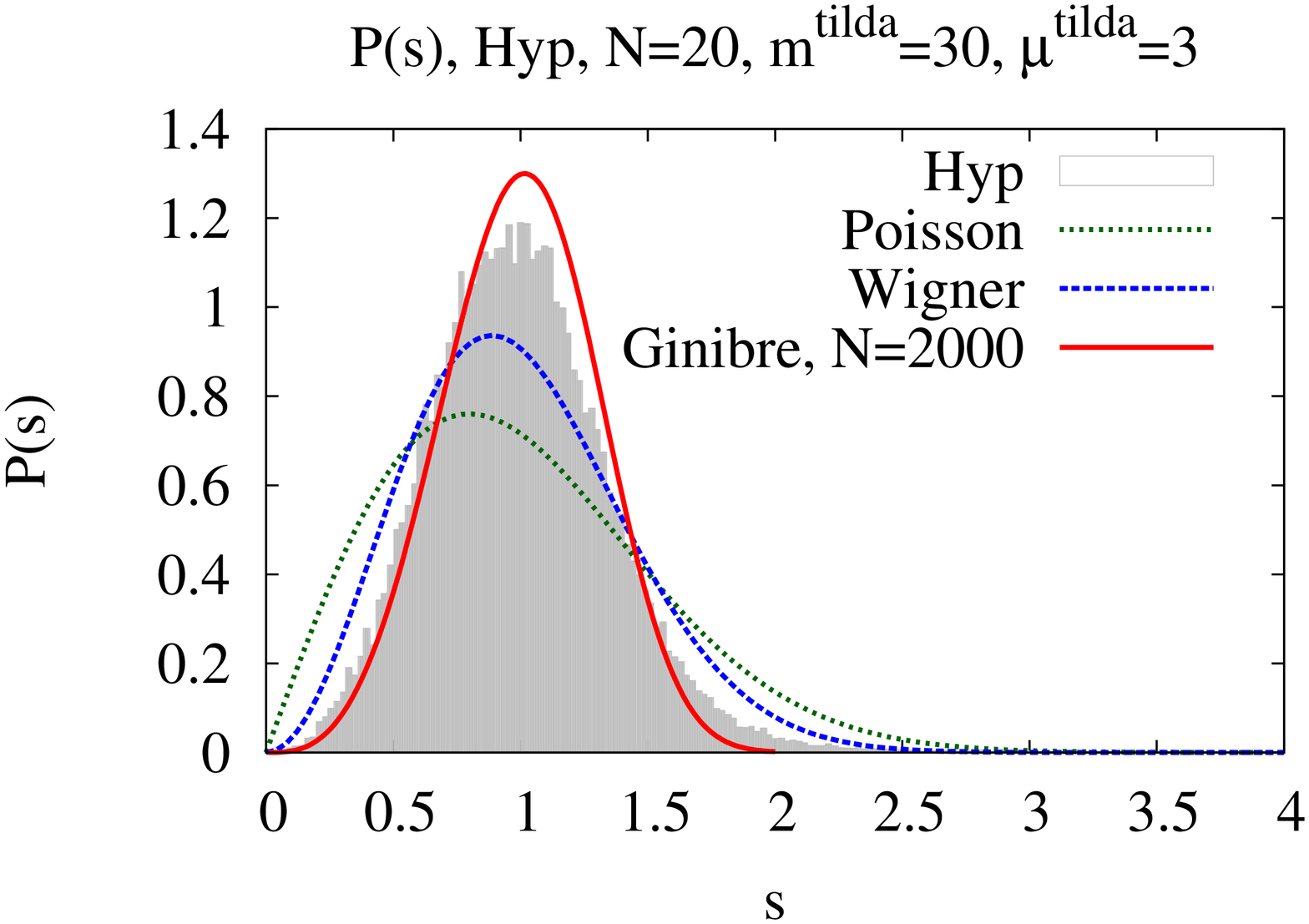}
  \end{center}
  \caption{
  The NNS distribution of 
  the Hyp notation at $N=20$, $\tilde{m}=30$, and $\tilde{\mu}=3$.
  }
  \label{fig:N20mu3m30H}
\end{figure}

We show the NNS distributions in the Hyp representation 
for ($\tilde{\mu}, \tilde{m})=(3,2)$ and (3,30) in Figs.~\ref{fig:N20mu3m2H} and \ref{fig:N20mu3m30H}, 
respectively.
The NNS distributions in those cases
are approximately consistent with each other, but slightly smaller than the Ginibre ensemble.
We do not find strong $\tilde{m}$ dependence of the NNS distribution 
in the Hyp representation.


\begin{figure}[htbp]
  \begin{center}
   \includegraphics[width=60mm, angle=270]{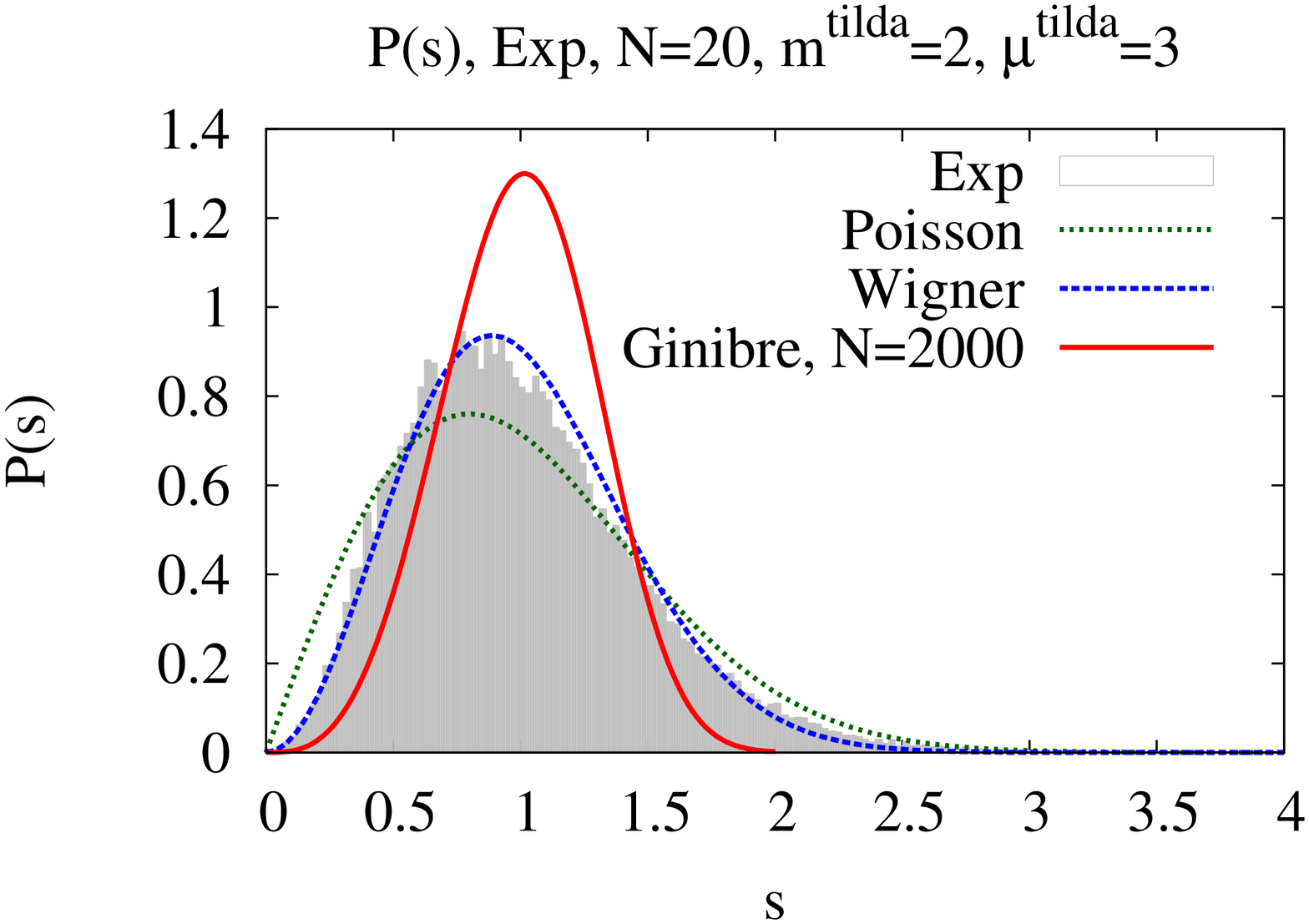}
  \end{center}
  \caption{
  The NNS distribution of 
  the Exp notation
  at $N=20$, $\tilde{m}=2$, and $\tilde{\mu}=3$.
  }
  \label{fig:N20mu3m2E}
  \begin{center}
   \includegraphics[width=60mm, angle=270]{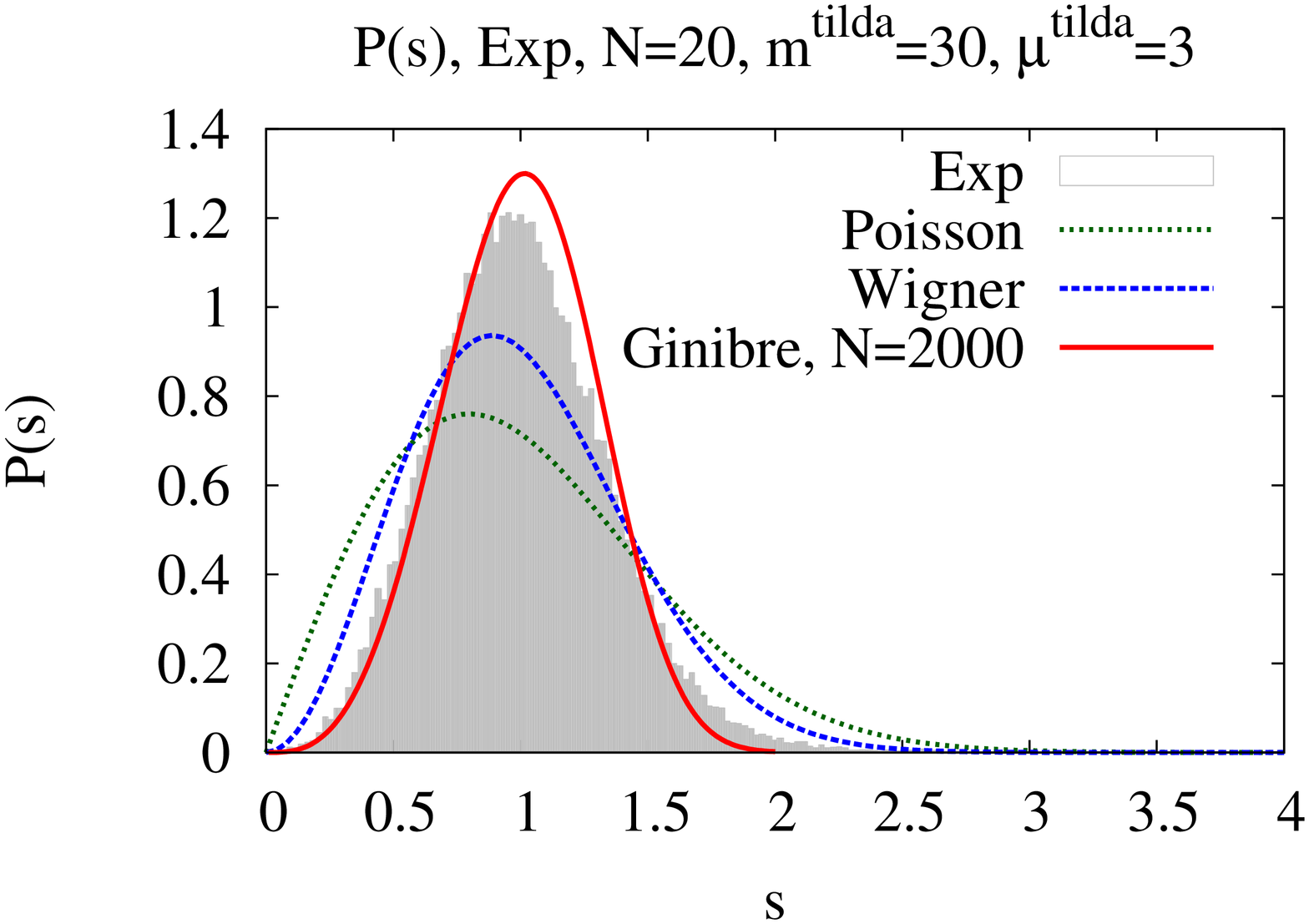}
  \end{center}
  \caption{
  The NNS distribution of 
  the Exp notation
  at $N=20$, $\tilde{m}=30$, and $\tilde{\mu}=3$.
  }
  \label{fig:N20mu3m30E}
\end{figure}




Next, we show the NNS distribution in the Exp representation
for ($\tilde{\mu}, \tilde{m})=(3,2)$ and (3,30) in Figs.~\ref{fig:N20mu3m2E} and \ref{fig:N20mu3m30E}, respectively.
We find that the NNS distributions depend on $\tilde{m}$.
At small $\tilde{m}$, the NNS distribution is close
to the Wigner surmise.
At large $\tilde{m}$, the NNS distribution approximately follows that in the Ginibre ensemble, and 
is almost consistent with that in the Hyp representation. 






\subsection{Discussion}

Now, we discuss the interpretation of our results and its implications.
We found that the Ginibre ensemble is 
favored at large $\tilde{m}$ in both representations.
This is physically reasonable, because the fermion determinant 
is approximated by the mass factor as $\det(\Slash{D}+m) \sim m^{2N}$.
There is no difference between the original theory and phase 
quenched theory, and both of them provide the same results.
In this case, the partition function is well described as 
the Gaussian form of the complex matrices $\Phi_{1,2}$, 
which is nothing but the Ginibre ensemble \cite{Ginibre:1965zz,Verbaarschot:2005rj}. 

On the other hand, we found that at small $\tilde{m}$ 
the Wigner surmise and Ginibre ensemble are favored in the Exp 
and Hyp representation, respectively. 
The ChRMT used in this work is independent of quark chemical potential $\mu$
~\cite{Osborn:2004rf,Bloch:2012bh}, which implies that the NNS distribution
at finite $\mu$ should be  the same with that at zero $\mu$.
As we have explained, 
the Wigner surmise is theoretically expected at $\mu = 0$.
Thus, we find that the NNS distribution shows physically expected 
behavior even in the CL method, if the Langevin simulation converges 
to the correct results.
The Ginibre ensemble in the Hyp representation is caused by 
unphysical broadening of the Dirac eigenvalue distribution 
due to the failure of the complex Langevin simulation.

Here, we should note that there is a subtlety in the application
of the CL method to the NNS distribution.
The argument to justify the CL method is given for holomorphic observables~\cite{Aarts:2009uq,Aarts:2011ax},
while it is unclear if the CL method can be justified
for non-holomorphic observables.
The NNS distribution is real, and therefore a non-holomorphic quantity.
It is not ensured that the NNS distribution obtained in the CL method agrees with that in the original theory.
Although the Dirac eigenvalue distribution is narrower for the correct
convergence case, the distribution more or less receives broadening caused by
the complexification, namely imaginary parts of the originally real variables.
In this sense, it is non-trivial that the CL method can reproduce the 
physical behavior of the NNS distribution. 

As we mentioned in the Introduction, we conjecture that the NNS distribution, 
 a universal quantity defined for the Dirac eigenvalues, 
can provide physical behavior even in the complex Langevin simulation. 
Although this is highly non-trivial statement due to the non-holomorphy of the 
NNS distribution, our numerical results support this conjecture.
Such a conjecture may be also inferred from an analogy between the
CL method and the Lefshetz thimble (LT) method. 
In the LT method, dominant critical points are expected to share the symmetry properties with an original integral contour
as shown in the case of one classical thimble contribution \cite{Cristoforetti:2012su}.
In the CL and LT methods, the critical points are located at the same points 
on the complex plane, and therefore dominant critical points in the CL method 
are also expected to share the symmetry properties of the original theory. 
If this conjecture holds, the universal quantities may provide a tool to 
understand the convergence properties of the CL simulations, which 
will be useful in the study of theories where exact results are not known 
such as in QCD.

\section{Summary} \label{Sec:Summary}
In this work, we have performed the first study of the nearest neighbor 
spacing (NNS) distributions of Dirac eigenvalues in the chiral random matrix theory (ChRMT) 
at non zero quark chemical potential using the complex Langevin (CL) method.
The ChRMT was described in two representations:
the hyperbolic and exponential forms of the chemical potential.
The polar coordinate was adopted in both cases for the description of dynamical variables. 
For small quark mass, the hyperbolic case converges to the wrong result 
and exponential case converges to the correct result, as shown in 
a previous study~\cite{Mollgaard:2014mga}.

We have calculated the NNS distribution for several values of the mass and
chemical potential using the unfolding procedure.
For large mass, the NNS distribution follows the Ginibre ensemble, which
implies that the real and imaginary part of the Dirac eigenvalues 
have the same order of magnitude. 
For small mass, we found the deviation between two representations.
The NNS distribution follows the Wigner surmise for the correctly converging case, 
while it follows the Ginibre ensemble when the simulation converges to the phase quenched result.
The Wigner surmise is physically reasonable according to the chemical potential 
independence of the ChRMT.
Thus, the NNS distribution shows the physical behavior even in the 
CL method if it converges to the correct result.

There is a subtlety as to whether the NNS distribution, which is
non-holomorphic, can be justified in the CL method. 
We speculate from the analogy between the CL and Lefshetz thimble 
methods that universal quantities determined from the 
properties such as symmetries can be maintained even in the CL method
if the configurations are correctly located around relevant critical points. 
If this holds, the universal quantities can be used to test the
convergence properties of the CL method. 
At least, our numerical result support this conjecture.
Of course, it is important to consider the theoretical justification 
of the conjecture and applications to other universal quantities and to other theories, 
which we leave for future studies.
Such a study will deepen our understanding of complexified theories, 
and may provide information about a convergence property in the 
CL simulation for the study of theories with the sign problem. 

\newpage

%
%
%

\acknowledgments
The authors would like to thank Falk Bruckmann, Sayantan Sharma, and Sinji Shimasaki for useful discussions.
T.~I.~thanks Yuta Yoshida for technical help.
T.~I.~ is supported by the Grant-in-Aid for the Japan Society for the Promotion of Science (JSPS) Fellows (No.~25-2059). 
K.~K.~ is supported by the Grant-in-Aid for the Japan Society for the Promotion of Science (JSPS) Fellows (No.~26-1717). 
K.~N.~ is supported by the JSPS Grants-in-Aid for Scientific Research (Kakenhi) Grants No.~26800154, 
and by MEXT SPIRE and JICFuS.
\appendix

\section{Drift terms}
\label{sec:driftterms}
The drift terms for the angular and radius variables in the
CL equations with the Hyp notation are gives as
\begin{widetext}
\ba
 - \frac{\partial S}{\partial \tM{1}{mn}} 
 & = - \Nf \left[  \GM{ij} \partial_{\tM{1}{mn}} ( X_{jk}Y_{ki}) \right] 
\nn \\
 & = - \Nf \cosh (\mu) \left[ \GM{ij} ( X_{jk}  \rM{1}{ik} e^{-i\tM{1}{ik}} \delta_{mi} \delta_{nk} 
  -  \delta_{mj} \delta_{n k} \rM{1}{j k} e^{i \tM{1}{jk}} Y_{ki}) \right] 
 \nn \\
  & = -\Nf \cosh (\mu) \left[ \GM{m j} X_{j n} \rM{1}{mn} e^{-i \tM{1}{mn}}
  -  \GM{i m} \rM{1}{mn} e^{i \tM{1}{mn}} Y_{n i}  \right]
  \nn \\
  & = - \Nf \cosh (\mu) \left[  (GX)_{mn} \rM{1}{mn} e^{-i \tM{1}{mn}}     - 
  (YG)_{nm}  \rM{1}{mn} e^{i \tM{1}{mn}}  \right]
  \nn \\
  &=  - \Nf \cosh (\mu) \rM{1}{mn} \left[  (GX)_{mn}  e^{-i \tM{1}{mn}}
   -  \left( (YG)^{t} \right)_{mn}  e^{i \tM{1}{mn}}  \right]  
 \ ,
 \end{align}
 \begin{align}
 - \frac{\partial S}{\partial \tM{2}{mn}} 
 & = - \Nf \left[  \GM{ij} \partial_{\tM{2}{mn}} (X_{jk}Y_{ki}) \right] 
 \nn \\
 & = - i \Nf  \sinh (\mu ) \left[  \GM{ij} ( X_{jk} \rM{2}{ki} e^{i \tM{2}{ki}} \delta_{km}\delta_{in} 
  -  \delta_{km} \delta_{jn} \rM{2}{kj} e^{-i\tM{2}{kj}}  Y_{ki}  ) \right]
 \nn \\
  & = - i \Nf  \sinh (\mu) \left[   \GM{n j} X_{j m} \rM{2}{mn} e^{i\tM{2}{mn}} 
   -\GM{in} \rM{2}{mn} e^{- i \tM{2}{mn}} Y_{mi} \right]
\nn \\
   & = - i \Nf  \sinh (\mu)   \rM{2}{mn} \left[ \PRTs{(GX)^t}{mn}  e^{i\tM{2}{mn}} 
   - \PRTs{YG}{mn} e^{- i \tM{2}{mn}}  \right]
   \ ,
\end{align}
\begin{align}
- \frac{\partial S}{\partial \rM{1}{mn}} 
 & = - 2 N \rM{1}{mn}  + \frac{1}{ \rM{1}{mn} }  
       - \Nf [ \GM{ij} \partial_{\rM{1}{mn}} X_{j k} Y_{k i} ]
\nn  \\
 & =- 2 N \rM{1}{mn}  + \frac{1}{ \rM{1}{mn} }  
           - i \Nf \cosh (\mu ) [ \GM{ij} \left(  e^{i\tM{1}{j k} } \delta_{jm} \delta_{kn} Y_{ki} 
           + X_{j k} e^{-i\tM{1}{i k}} \delta_{m i} \delta_{n k} \right)]
      \nn     \\
 & =- 2 N \rM{1}{mn}  + \frac{1}{ \rM{1}{mn} }  
           - i \Nf \cosh (\mu ) [ \GM{i m}   e^{i\tM{1}{m n} } Y_{n i} 
           + \GM{m j} X_{j n} e^{-i\tM{1}{m n}}  ]
    \nn       \\
 & =- 2 N \rM{1}{mn}  + \frac{1}{ \rM{1}{mn} }  
           - i \Nf \cosh (\mu ) [ \PRTs{(YG)^t}{mn}   e^{i\tM{1}{m n} } 
           + \PRTs{GX}{mn} e^{-i\tM{1}{m n}}  ]
       \ , 
   \end{align}
   \begin{align}
- \frac{\partial S}{\partial \rM{2}{mn}} 
 & = - 2 N \rM{2}{mn}  + \frac{1}{ \rM{2}{mn} }  
       - \Nf [ \GM{ij} \partial_{\rM{2}{mn}} X_{j k} Y_{k i} ]
  \nn \\
 & =-  2N \rM{2}{mn}  + \frac{1}{ \rM{2}{mn} }  
           -  \Nf \sinh (\mu ) [ \GM{ij} \left(  e^{ - i\tM{2}{k j} } \delta_{k m} \delta_{j n} Y_{k i} 
    \nn       + X_{j k} e^{ i\tM{2}{k i}} \delta_{m k} \delta_{n i} \right)]
           \\
 & =- 2 N \rM{2}{mn}  + \frac{1}{ \rM{2}{mn} }  
           -  \Nf \sinh (\mu ) [ \GM{i n}   e^{- i\tM{2}{m n} } Y_{m i} 
           + \GM{n j} X_{j m} e^{ i\tM{2}{m n}}  ]
      \nn     \\
 & =- 2 N \rM{2}{mn}  + \frac{1}{ \rM{2}{mn} }  
           -  \Nf \sinh (\mu ) [ \PRTs{YG}{mn}   e^{- i\tM{2}{m n} } 
           + \PRTs{( GX )^t }{mn} e^{ i\tM{2}{m n}}  ]
           \ ,
 \end{align}
\end{widetext}
where 
we use a formula, $(\det X)^{'} = \det X \  \Tr (X^{-1} X^{'} )$. 


\section{Results on the nearest neighbor distributions with two fitting function} \label{Sec:Fit}

\begin{figure}[htbp]
   \includegraphics[width=60mm, angle=270]{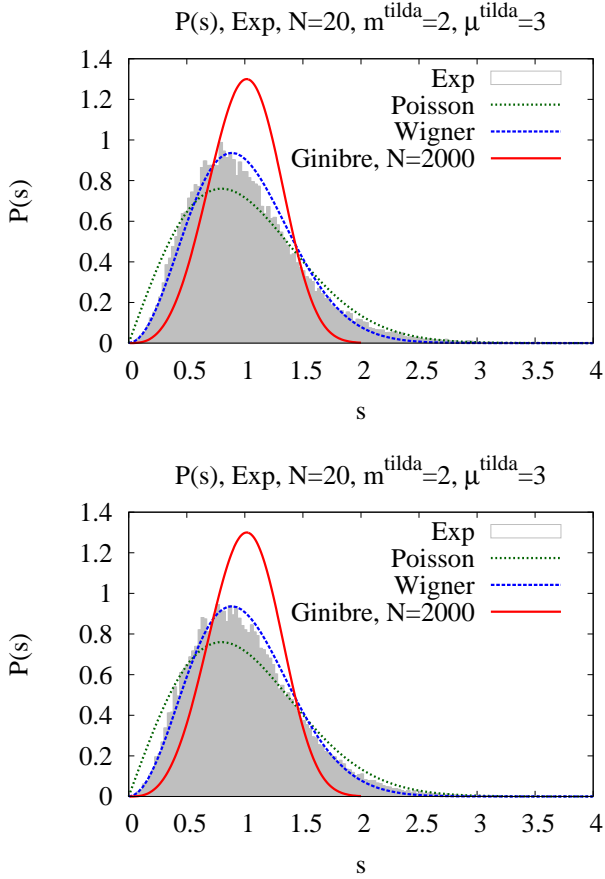}
   \includegraphics[width=60mm, angle=270]{Ps-tot-exp-N20-mass2-mu3.eps}
  \caption{
  The NNS distribution
  at $N=20$, $\tilde{m}=2$, and $\tilde{\mu}=3$
  in Exp
 with a cubic fitting function in the top panel and that with a
 quintic fitting function in the bottom panel.
  \label{fig:Ps-Exp-N20mu3m2}
  }
\end{figure}

\begin{figure}[htbp]
   \includegraphics[width=60mm, angle=270]{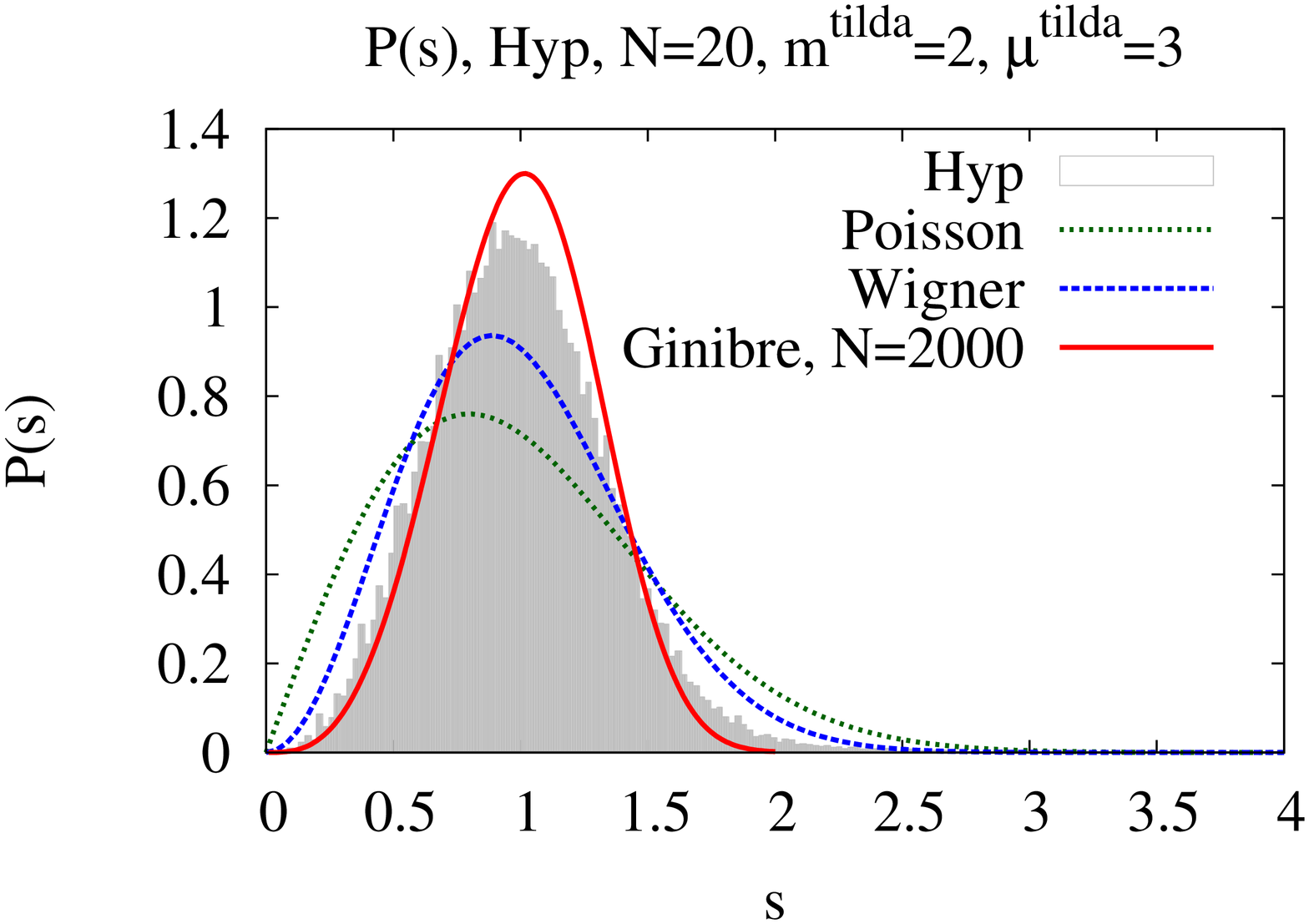}
   \includegraphics[width=60mm, angle=270]{Ps-tot-hyp-N20-mass2-mu3.eps}
  \caption{
  The NNS distribution
  at $N=20$, $\tilde{m}=2$, and $\tilde{\mu}=3$
  in Hyp with a cubic fitting function in the top panel and that with a
 quintic fitting function in the bottom panel.
  \label{fig:Ps-Hyp-N20mu3m2}
  }
\end{figure}

In Figs.~\ref{fig:Ps-Exp-N20mu3m2} and \ref{fig:Ps-Hyp-N20mu3m2},
the NNS distributions in Hyp and Exp with two fitting functions.
The dependence of the fitting functions seems to be small for the NNS distributions.
Then, we adopt a quintic function to obtain the NNS distributions in this paper.

\bibliography{ref,ref_addedbyKN}

\end{document}